\documentclass[aip,jap,amssymb,reprint]{revtex4-1}

\usepackage{graphicx}
\usepackage{dcolumn}
\usepackage{bm}

\newcommand{\ee}[1]{\ensuremath{\times 10^{#1}}}	
\newcommand{\Ei}[0]{\ensuremath{E_{\rm i}}}				
\newcommand{\Eh}[0]{\ensuremath{E_{\rm h}}}				
\newcommand{\Eib}[0]{\ensuremath{E_{\rm ib}}}				
\newcommand{\Em}[0]{\ensuremath{E_{\rm m}}}				
\newcommand{\Pm}[0]{\ensuremath{P_{\rm m}}}				
\newcommand{\Pib}[0]{\ensuremath{P_{\rm ib}}}				
\newcommand{\Fdet}[0]{\ensuremath{F_{\rm det}}}				

\begin{document}

\title[Cryogenic micro-calorimeters for mass spectrometry]{Cryogenic
micro-calorimeters for mass spectrometric identification\\
of neutral molecules and molecular fragments}

\author{O Novotn\'y}
\email{oldrich.novotny@mpi-hd.mpg.de}
\affiliation{Max-Planck-Institut f\"{u}r Kernphysik, D-69117 Heidelberg,
Germany}
\affiliation{Columbia Astrophysics Laboratory, Columbia University, New
York, NY 10027, USA}
\author{S Allgeier}
\author{C~Enss}
\author{A~Fleischmann}
\author{L~Gamer}
\author{D~Hengstler}
\author{S~Kempf}
\affiliation{Kirchhoff Institute for Physics, Heidelberg University, D-69120
Heidelberg, Germany}
\author{C~Krantz}
\affiliation{Max-Planck-Institut f\"{u}r Kernphysik, D-69117 Heidelberg,
Germany}
\author{A~Pabinger}
\author{C~Pies}
\affiliation{Kirchhoff Institute for Physics, Heidelberg University, D-69120
Heidelberg, Germany}
\author{D~W~Savin}
\affiliation{Columbia Astrophysics Laboratory, Columbia
University, New York, NY 10027, USA}
\author{D~Schwalm}
\affiliation{Max-Planck-Institut f\"{u}r Kernphysik, D-69117 Heidelberg,
Germany}
\affiliation{Faculty of Physics, Weizmann Institute of Science, Rehovot 76100,
Israel}
\author{A~Wolf}
\affiliation{Max-Planck-Institut f\"{u}r Kernphysik, D-69117 Heidelberg,
Germany}

\date{\today}

\begin{abstract}
We have systematically investigated the energy resolution of a magnetic
micro-calorimeter (MMC) for atomic and molecular projectiles at impact energies
ranging from $E\approx13$ to 150~keV. For atoms we obtained absolute energy
resolutions down to $\Delta E \approx 120$~eV and relative energy resolutions
down to $\Delta E/E\approx10^{-3}$. We also studied in detail the MMC
energy-response function to molecular projectiles of up to mass 56~u. We have
demonstrated the capability of identifying neutral fragmentation products of
these molecules by calorimetric mass spectrometry. We have modeled the MMC
energy-response function for molecular projectiles and conclude that
backscattering is the dominant source of the energy spread at the impact
energies investigated. We have successfully demonstrated the use of a detector
absorber coating to suppress such spreads. We briefly outline the use of MMC
detectors in experiments on gas-phase collision reactions with neutral products.
Our findings are of general interest for mass spectrometric techniques,
particularly for those desiring to make neutral-particle mass measurements.
\end{abstract}


\maketitle

\section{Mass spectrometry for multi-keV neutral particles}\label{sec:intro}
Mass spectrometry is one of the most important analytical techniques used in
both fundamental sciences and industry applications. The diverse fields of use
include protein and genome characterization in molecular biology
\cite{Aebersold:Nature:2003,Gstaiger:Nature:2009}, determination of fundamental
constants \cite{Sturm:Nature:2014}, atmosphere composition characterization for
Earth \cite{deGouw:MSR:2007} and other solar system bodies
\cite{Niemann:Nature:2005}, isotope dating \cite{Li:Nature:1989}, drug discovery
\cite{Rossi:book:2001}, etc. Most molecular mass-spectrometry methods employ
electromagnetic fields to determine mass-to-charge ratios of ionized particles
\cite{Gross:book:2011}. Two major limitations arise from this approach. First,
the mass-to-charge ratios may be ambiguous, e.g., singly charged monomers are
indistinguishable from doubly charged dimers. Second, neutral fragments cannot
be analyzed directly and a ``neutral loss'' must be accepted. To circumvent
these limitations, the neutrals studied can be ionized and then analyzed in a
standard way using electromagnetic fields \cite{Wesdemiotis:CR:1987}. Although 
``gentle'' ionization methods exist nowadays (e.g., electrospray ionization
\cite{Fenn:Science:1989}), the ionization often results in further fragmentation
and in multiple ionization. Each of these processes can lead to a significant
increase of complexity in the measured mass-to-charge spectrum
\cite{Mann:ARB:2001}. Unfolding such data into a mass spectrum of the original
sample may be complicated \cite{Sharon:JASMS:2010} and often only qualitative
information can be retrieved \cite{Steen:MCP:2006}.

Calorimetry represents an alternative  approach to measure the masses of neutral
particles. Calorimetric detection can be based on ionization, such as in the
surface barrier or scintillation detectors  applied mainly in nuclear and
particle physics \cite{Leo:book:1994}, or on thermal effects, as we discuss
below. In all of these approaches, the calorimetric signal is proportional to
the energy set free by the stopping of an incident particle, mostly given by its
initial kinetic energy. Masses can be deduced from the magnitude of the kinetic
energy if the incident velocity is known.

In molecular physics, calorimetric detection has been applied in particular  to
gas-phase collision studies using fast particle beams. For those studies many of
the reaction products are neutral and their detection and mass measurement is
essential in order to determine reaction cross sections and fragmentation
branching ratios. A range of studies have been devoted to collision-induced
dissociation (CID) of molecular and cluster ions in a fast beam, interacting
with a stationary gas target (see, e.g.,
\cite{Chabot:NIMA:2002,Abdoul:IJMS:2014} for recent work). Extensive work has
also been  directed to the study of dissociative recombination (DR) of molecular
ions with electrons, which for singly charged ions produces exclusively neutral
fragments \cite{Florescu-mitchell:PR:2006}. DR is closely related to the process
of electron capture dissociation (ECD), which is widely applied in the field of
mass spectrometry for sequencing multiply-charged biomolecular cations
\cite{Zubarev:JACS:1998, Syrstad:JASMS:2005}.

The DR reaction on singly charged ions was studied in particular with fast ion
beams in a storage ring, utilizing a merged electron-ion beams configuration
\cite{Phaneuf:RPP:1999} where the electron-ion interaction energy could be tuned
to small (sub-eV) values. Considering this reaction on a model molecule
ABCD$^+$, the neutral fragmentation channels may include, e.g.,   
\begin{eqnarray}
 {\rm ABCD^+}+e^- & \rightarrow & \rm A +BCD\\
 & \rightarrow & \rm AB + CD \\
 & \rightarrow & \rm A + B + C + D.\label{eq:DR}
\end{eqnarray}
In a fast ion beam, the parent ion velocity (corresponding to multi-keV or MeV
ion beam energies) is typically much larger than the relative velocities that
the fragments gain in the dissociation process (from the typical kinetic energy
release of $\lesssim 10$~eV). As a result, the DR products arising from the beam
overlap region have velocities close to that of the parent ions. After
separating the ion beam by means of magnetic or electrostatic fields the neutral
DR products are collected by a kinetic-energy-sensitive detector. From the known
parent ion velocity, and from the measured fragment kinetic energies, the
fragment masses can in general be easily assigned. This method has been
successfully used in combination with magnetic heavy ion storage rings
\cite{Larsson:ARPC:1997,Al-khalili:PRA:2003,Wolf:JPCS:2011,Novotny:APJ:2014}
operating at $\sim$~MeV ion beam energies and utilizing surface-barrier
detectors (SBDs)~\cite{Buhr:PRA:2010}. The approach, however, is restricted to
high beam energies as SBD detectors have a thin, but significant, insensitive
surface layer which cannot be penetrated by low-energy particles
\cite{Steinbauer:NIMB:1994}. Moreover, even at MeV beam energies the energy
resolution of SBDs is not sufficient to identify heavy neutral molecular
fragments with 1~u resolution~\cite{vigren:ApJ:2010}. 

Stored ion beams of lower energy have been attracting significant attention for
some time now \cite{Andersen:JPB:2004}. One emphasis of these studies has been
on complex molecular ions. In order to optimize the storage times of such ions
and minimize the influence of thermal radiation, ion  beam traps and storage
rings involving cryogenic cooling of the enclosing experimental vacuum chamber
have recently been commissioned or are currently under development. In these
electrostatic storage devices \cite{Lange:RSI:2010, Thomas:RSI:2011,
Nakano:JPCS:2012, Schmidt:RSI:2013} the experimental vacuum chamber is cooled to
$\sim 10$~K. The beam energy for stored singly charged ions typically ranges up
to some $E\sim30$~keV. Higher ion beam energies, up to 300~keV for singly
charged ions, will be used in the cryogenic storage ring (CSR)
\cite{Von_Hahn:NIMB:2011,Krantz:JPCS:2011} at the Max-Planck-Institute for
Nuclear Physics in Heidelberg, Germany. This ring is also being equipped for
electron-ion merged beams experiments \cite{Krantz:JPCS:2011} to allow for DR
experiments on complex molecular ions. It is this CSR development that mainly
motivates the present search for alternative calorimetric detection techniques
for molecular particles.

A promising approach for solving the needs for low kinetic energy measurements
is the use of cryo-detectors \cite{Enss:book:2005}, such as Superconducting
Tunnel Junction detectors (STJ) or micro-calorimeters. In these detectors the
incident particle excites various types of energy states in the absorbing bulk.
Depending on the detector type, this excitation can be transformed into
electrical signal. For example, in micro-calorimeters the particle kinetic
energy is transformed into heat and a sensitive temperature monitoring then
provides the measure of the kinetic energy deposited. For cryo-detectors,
operation at low temperatures ($\lesssim 1$~K) is needed to limit thermal noise
and thus resolve the measured low energies. The advantages of cryo-detectors
were recognized as early as 1935 \cite{Simon:Nature:1935}. Their broader use was
first motivated by astrophysical applications, e.g., in dark matter searches
\cite{Cabrera:PRL:1985}, and for high-precision X-ray spectroscopy
\cite{Twerenbold:EL:1986}. 

In the 1990s the advantages of cryo-detectors were also recognized for classical
mass spectrometry \cite{Twerenbold:NIMA:1996,Hilton:Nature:1998}. Aside from the
benefits explained above, an additional motivation for using cryo-detectors was
their ability to resolve very slow massive molecules where standard microchannel
plate detectors fail. Compared to microchannel plates, calorimetric detectors
also offer a single-particle detection efficiency that could, in principle,
reach values very close to unity. On the other hand, the aspect ratio of the
channels (typically about 0.7) limits the efficiency of microchannel plates for
detecting single-particle impacts to significantly below unity for many
applications. Various types of cryo-detectors have been implemented in previous
mass spectrometric studies, mainly in time-of-flight mass spectrometry (TOF-MS)
\cite{Hilton:Nature:1998, Frank:RCMS:1996,Westmacott:RCMS:2000,Suzuki:PCS:2009}.
Recently, and independently of our own attempts, a cryogenic STJ detector has
been employed in a tandem mass spectrometer \cite{Ohkubo:IJMS:2011}  following
an approach similar to the calorimetric mass determination discussed above for
gas-phase molecular collision and ion storage ring studies. 

When used for X-ray spectroscopy \cite{Porst:JLTP:2014} cryo-detectors can
achieve an excellent relative energy resolution of $\Delta E/E\approx10^{-4}$
and an absolute resolution down to $\Delta E\approx 1$~eV. For detecting massive
particles, however, the resolution degrades by orders of magnitude. In heavy-ion
experiments performed at MeV and GeV energy, resolutions of not better than
$\sim
10^{-3}$ have been reached \cite{Egelhof:book:2005}. For $\alpha$-spectrometry
studies researchers measured $\Delta E/E\approx5\times10^{-4}$ at $E=5.5$~MeV
\cite{Horansky:JAP:2010,Ranitzsch:NIMA:2011,Yoo:JLTP:2012}. However, energy
resolutions of only $\sim10^{-1}$ were reported for detecting atoms and
molecules at energies of $\lesssim10$~keV
\cite{Hilton:Nature:1998,Frank:RCMS:1996,Westmacott:RCMS:2000,
Ohkubo:IJMS:2011,Tomita:APL:2007,Ohkubo:TAS:2014}, i.e., in the regime most
relevant for mass spectrometry. Only a few studies have attempted to understand
this issue. Andersen \cite{Andersen:NIMB:1986} theoretically predicted the
cryogenic calorimeter energy resolution for heavy ions. In his calculations he
included various projectile energy loss processes in the detector absorber, such
as backscattering and Frenkel-pair formation, and estimated the amount of energy
which is not converted into heat and thus cannot be detected. The statistical
nature of the losses then defines the broadening in the acquired energy spectra.
Quantitatively, Andersen's predictions for various detector types match to
within an order of magnitude the energy resolutions measured later at MeV to GeV
energies \cite{Egelhof:book:2005}. More recently Horansky et al.\ have
numerically modeled the micro-calorimeter energy spectra from 5.5~MeV
$\alpha$-particles \cite{Horansky:JAP:2010} and have concluded that the dominant
spectral broadening is due to lattice damages. Their model underestimates the
energy spread  by a factor of $\sim$~2 compared to the measurements. We are not
aware of any study attempting to theoretically describe the energy resolution of
a cryo-detectors for massive particles at $E<100$~keV energies. Similarly, to
the best of our knowledge there are no systematic experimental studies of the
energy resolution for cryo-detectors as a function of projectile energy and type
for sub-MeV energies. In the energy range from $\sim 10$~keV to $5.5$~MeV, even
demonstration measurements are missing. 

We therefore believe that the first step for understanding and eventually
improving the cryo-detector
energy resolution for mass spectrometry is a systematic study of the detector
energy-response function for atomic and simple molecular projectiles at
sub-MeV energies. To this end we have experimentally studied the energy
resolution of a micro-calorimeter detector for atomic and molecular ions and
neutrals at energies from $\sim 13$ to 150~keV. The rest of the paper
is organized as follows: In Section~\ref{sec:MMC} we briefly introduce the
specific class of micro-calorimeters employed in our experiments, namely the
metallic magnetic calorimeter (MMC). In Section~\ref{sec:exp} our experimental
setup is described. In Section~\ref{sec:results} we present the acquired energy
spectra and a corresponding model resolving the various energy loss processes.
Based on this model, we also demonstrate the importance of detector absorber
materials. We summarize and present our future plans in
Section~\ref{sec:sumout}.

\section{Metallic magnetic calorimeter (MMC) detectors}\label{sec:MMC}
\subsection{Detection principle}
Micro-calorimeters are cryo-detectors typically operating  at $\lesssim 100$~mK
\cite{Enss:book:2005}. In this temperature regime the kinetic energy of a
detected particle heats up the detector absorber such that the increase of the
absorber temperature is proportional to the original projectile kinetic energy.
The response is linear to a high degree. In order to monitor the absorber
temperature, various sensor types can be used. MMCs are operated with 
paramagnetic sensors whose magnetization at low temperatures is a monotonic
function of temperature. The magnetization change is measured with a
superconducting coil inductively coupled to a superconducting quantum
interference device (SQUID). A weak thermal link between the absorber and a cold
thermal bath ensures that  the absorber cools back to the bath temperature after
the energy readout. A detailed description of the MMC operation principle and
micro-fabrication can be found in
\cite{Enss::JLTP:2000,Fleischmann:book:2005,Fleischmann:AIP:2009}.

\subsection{Energy resolution}\label{sec:MMCres}
In general the energy resolution of an MMC detector setup is given by the
thermal noise in the detector absorber and sensor, by the electronic noise in
the signal processing system, and by the processes transforming the kinetic
energy of the detected particle into absorbed heat. The thermal and electronic
noise levels do not depend on the nature and energy of the detected particles
and thus define the intrinsic energy resolution of the detector. For MMCs the
intrinsic energy spread is typically $\lesssim 100$~eV (full width at half
maximum, FWHM) and under very favourable conditions energy spreads of $\lesssim
1.6$~eV have been achieved \cite{Porst:JLTP:2014, Fleischmann:prep:2015}.

The kinetic energy transformation into heat is very efficient for X-rays. There
the energy absorption process is dominated by the photoelectric effect, followed
by the photoelectron efficiently heating the electron gas and subsequent energy
transfer to the phonons \cite{Enss::JLTP:2000}. Only a negligible part of the
energy is lost, e.g., by kinetic energy of secondary electrons escaping the
surface of the absorber. In contrast to detecting energetic photons, stopping
atomic and molecular projectiles in the absorber may lead to additional
processes, such as backscattering of the projectiles, sputtering of absorber
material, and absorber lattice defect formation. All of these non-thermal
processes remove part of the projectile kinetic energy before it can be
converted into heat in the absorber and measured by the temperature sensor.
Moreover, the statistical nature of these processes results in a spread of the
detected energies. Thus in general the energy resolution of an MMC detector (and
cryo-detectors in general) depends on the absorber material as well as on the
nature and energy of the impacting particles.

\section{Experiment for determining MMC energy resolution}\label{sec:exp}

In order to experimentally determine the energy resolution of an MMC detector we
have measured kinetic energies $\Em$ of single ions from a nearly monoenergetic,
mass-to-charge-selected ion beam. The accumulated single-particle calorimetric
signals, after suitable calibration to yield $\Em$, were used to derive energy
spectra $\Pm(\Em,\Eib,M)$ which, in turn, were used to determine the
corresponding energy resolutions. The measured spectra are labeled by the
nominal ion beam energy $\Eib$ and by the ion mass $M$ (since the systems
studied are simple enough to uniquely specify the ion type from its mass).

\subsection{Ion beam production} \label{sec:ionprod}
Ion beams were generated using the high-voltage ion beam platform at the Max
Planck Institute for Nuclear Physics in Heidelberg, Germany
\cite{Krantz:JPCS:2011}. The various singly charged molecular cations  were
produced in a gas discharge from a Penning ion source. While still on the
platform, they were accelerated  in a first stage by $\sim 13$~kV  and then
filtered  by their mass-to-charge ratio in a $90^\circ$ dipole magnet. After
passing through a set of apertures the ions were further accelerated in a second
stage by the potential applied to the platform relative to ground. By adjusting
the platform voltage, the  final ion energies were varied in the range from
$\Eib\approx 13$ to 150~keV. After leaving the platform the ion
beam was focused and directed by an electrostatic quadrupole and a set of
magnetic deflectors onto the detector setup  $\sim 10$~m from the platform
exit. 

The nominal beam energy $\Eib$ was determined as a sum of the two involved
acceleration voltages used. The estimated systematic uncertainties in the
$\Eib$ scaling and offset are $\pm0.5$\% and $\pm300$~eV,
respectively. Here and throughout all uncertainties are quoted at an estimated
$1\sigma$ statistical confidence level.

Some of the tests turned out to be sensitive to instabilities in the
acceleration voltages, resulting in a spread of individual ion energies $\Ei$.
We approximate the ion energy distribution $P_{\rm ib}(\Ei, \Eib)$ by a
symmetrical energy spread function peaking at $\Eib$ with a FWHM of
$\Delta{\rm_{ib}}$.

The voltage for the first-stage acceleration on the high-voltage platform was
stable to within $\approx 5$~V. For the second-stage acceleration by the
high-voltage platform potential, two different voltage sources have been used.
The first voltage source, hereafter denoted as A, was limited to 40~kV and
was used for most measurements with $\Eib \lesssim 53$~keV. This
source was stable to within $\approx 4$~V. The spreads from the first- and
second-stage acceleration, added in quadrature,  result in a total ion energy
spread of $\Delta{\rm_{ib}^A}\approx 6$~eV. As will be shown in
Section~\ref{sec:resato}, this ion energy spread is negligibly small in
comparison to the detector resolution.

For $\Eib >53$~keV a less stable voltage source B was employed in the
second-stage acceleration. The voltage fluctuations here dominate over
those from the first-stage acceleration.
Direct determination of the voltage spread was difficult and therefore we
derived it from the difference in peak widths in the detected energy spectra
$\Pm^{\rm A}$ and $\Pm^{\rm B}$ acquired with the same ion type, the same
$\Eib$, but with voltage sources A and B, respectively. As shown below in
Section~\ref{sec:resato}, the spectra for atomic projectiles are dominated by a
single Gaussian-like peak. Thus, by approximating $P_{\rm ib}(\Ei,\Eib)$ with a
normal distribution we estimated the ion beam energy spread for the voltage
source B as 
\begin{equation} 
\Delta_{\rm ib}^{\rm B} \approx \sqrt{(\Delta_{\rm
m}^{\rm B})^2 - (\Delta_{\rm m}^{\rm A})^2,} \label{eq:dib} 
\end{equation} 
where
$\Delta_{\rm m}^{\rm A}$ and $\Delta_{\rm m}^{\rm B}$ are the FWHMs of $\Pm^{\rm
A}$ and $\Pm^{\rm B}$, respectively. We applied the procedure separately for the
H$^+$ and Ar$^+$ data acquired at $\sim 53$~keV and obtained values of
$\Delta_{\rm ib}^{\rm B}$ which are equal to within their statistical
uncertainties. The error-weighted average of these two results is
$\Delta{\rm_{ib}^B}=320$~eV. A conservative estimate of the total systematic
error of this procedure is $\pm100$~eV, which exceeds the statistical error by a
factor of $\sim 2$. In Section~\ref{sec:resato} we discuss how we are able to
improve our estimate for the systematic error.

\subsection{Detector setup}
The detector setup was developed at the Kirchhoff Institute for Physics of the
University of Heidelberg, Germany. The MMC detector itself was derived from the
well-established \emph{maXs-200} design, which was originally developed for
X-ray spectroscopy \cite{Pies:JLTP:2012}. For the present measurements we have
used only two detector pixels, each having a $0.45\times2.0$~mm$^2$ sensitive
area. The absorber for one of the pixels was made of a $50~\mu\rm m$ thick gold
layer. If not explicitly mentioned otherwise, the  data presented below have
been obtained from this pixel. Selected measurements were performed with the
second pixel where the $50~\mu\rm m$  thick gold absorber was coated with an
additional 180~nm  thick layer of aluminum. In both cases incident atoms and
molecules  were stopped in the uppermost $\sim 150$~nm layer of material, as
verified by calculations using the SRIM software (``Stopping and Range of Ions
in Matter'' \cite{Ziegler:NIMB:2010}). Hence, the two detector pixels correspond
to stopping in gold and in aluminum, respectively. The absorber itself  has full
contact to a 2.4~$\mu$m thick thermal sensor made of a dilute gold-erbium alloy,
mixed at a mole fraction of [Er]/[Au+Er]=885~ppm. The erbium was depleted in
$^{167}$Er, which has a large hyperfine splitting and would otherwise add a
large heat capacity at the operational temperature. A detailed description of
the \emph{maXs-200} microfabrication and performance has been given previously
\cite{Pies:JLTP:2012,Pies:AIPCP:2009}.

The $\sim 10$~mK working temperature of our MMC detector was maintained with a
dilution refrigerator. To minimize the heat flux from the 300~K black-body
radiation of the ion beamline onto the detector, a set of radiation shields at
800~mK, 4~K, and 50~K surrounded the detector. Traveling along the ion beam
trajectory, the beam encountered a set of sequential $\sim1$~cm apertures at
50~K followed by a single 4~K foil with $\sim 4~\mu$m-diameter pinholes to limit
the radiation flux onto the detector head to below 1~nW.  The pinholes were
located so that there was at most one in front of each detector pixel. The pixel
used in the analysis was selected by connecting the readout electronics
described below. The rate of detected events was maximized before each data run
by slightly steering the ion beam.

\subsection{Data acquisition and processing}\label{sec:acqproc}
For each combination of ion species and ion beam
energy we accumulated between 500 and 2000 detection events. For
each event the electronic pulse from the thermal sensor was read out via the
SQUID electronics, amplified, and digitized.  The individual pulses were
analyzed to derive the detected energies $\Em$ and thus also the energy 
distribution $\Pm(\Em,\Eib,M)$.

The electronic signal of the detector reflects the temperature of the magnetic
moments of the paramagnetic sensor, which closely follows the absorber
temperature. The temporal evolution of the signal is determined  by the thermal
links between the absorber, the sensor, and the thermal bath. In our setup a
typical pulse reached its maximum amplitude within $\sim30$~$\mu$s after the
detection. The subsequent cooling results in an exponential signal decay with
a time constant of $\sim 5$~ms. In our experiments the energies detected were
small enough so that the sensor heat capacity can be approximated by a constant
over the whole range of sensor temperatures occurring during the detection
event. Additionally, the detector-operation temperature range was chosen so that
the sensor magnetization was proportional to the sensor temperature. As a
result, the shapes of the acquired electronic pulses were independent of the
detected energy. Moreover, to a very good approximation the pulse amplitudes
were linearly proportional to the heat released in the absorber. To obtain the
detected
energies on a relative scale we fitted the pulse amplitudes using a fixed pulse
shape. Specifically we used the advanced fitting procedure based on optimal
filtering, also called ``matched filter''
\cite{Szymkowiak:JLTP:1993,Fleischmann:RSI:2003}.

To put the detected kinetic energies on an absolute scale we additionally
detected photons from an $^{241}\rm\!Am\ \gamma$-source placed at $\sim 15$~cm
distance from the detector. The photon energy from such a source peaks sharply
at 59.5409~keV \cite{Helmer:NIMA:2000}. In our setup we have determined the
$\gamma$ energies with $\sim 15$~eV precision, which corresponds to $\sim
0.25$\% uncertainty in the scale factor of the energy calibration. The
non-linearity of the energy scale is expected to be less than 0.1\% at 60~keV
\cite{Keller:bachelor:2014}.

\subsection{Detector energy-response function} \label{sec:DetRespFnc}
The detector energy-response function relates the impacting ion kinetic energy
$\Ei$ to the measured energy $\Em$. As explained in Section~\ref{sec:MMCres},
the energy-response function of an MMC detector setup is determined by
statistical processes. First,  thermal processes in the absorber transform the
kinetic energy $\Ei$ into heat $\Eh$ with a statistical distribution $\Fdet^{\rm
TP}(\Eh,\Ei,M)$. This process may depend on the type of incident particles
denoted by $M$. Second, the thermal noise in the detector and the various noises
in the readout system convert $\Eh$ into the detected energy $\Em$. The response
function corresponding to this second transform, $\Fdet^{\rm 0}(\Em,\Eh)$,
reflects the intrinsic detector resolution. The overall detector response
function is then a convolution of the two spreads in the $\Eh$-domain:
\begin{equation}
 \Fdet(\Em,\Ei,M) = \Fdet^{\rm
0}(\Em,\Eh) \otimes \Fdet^{\rm TP}(\Eh,\Ei,M).\label{eq:FdetConv}
\end{equation}
We show below that the detector energy-response function $\Fdet$ can have a
complex shape, but is still often dominated by only a single peak. When that
is the case we can define the detector resolution $\Delta_{\rm det}(\Ei,M)$ as
the FWHM of $\Fdet(\Em)$ for a given $\Ei$ and projectile $M$. The relative
resolution is then defined as $\Delta_{\rm det}(\Ei,M)/\Ei$.

We have determined the intrinsic-resolution function of our detector setup,
$\Fdet^{\rm 0}$, by applying our standard event analysis procedure (as described
above) on the time-dependent electronic signal acquired between the detection
events. Fluctuations of this baseline signal arise from the same instabilities
and noise sources in the detector and  the data acquisition system  that also
act on the particle-generated pulses. As explained above, the detector signal
scales linearly with the energy input. Thus the pulses derived from this
baseline fluctuation reflect the intrinsic noise of the detection system and the
distribution of their peak amplitudes can be used as a measure for the intrinsic
energy resolution of our detector setup. The inferred $\Fdet^{\rm 0}$ displayed
a Gaussian shape centered at 0~eV with a FWHM of $\Delta{\rm_{det}^0}=107\pm
5$~eV.

\section{Results and discussion}\label{sec:results}
We have acquired MMC energy spectra $\Pm(\Em)$ for various singly charged atomic
and molecular ions in the energy range from $\Eib\approx13$ to 150~keV. We first
present our measurements with atomic ions in Section~\ref{sec:resato}. In
Section \ref{sec:masspec} we demonstrate the mass spectrometric capabilities of
the MMC detector using molecular fragmentation measurements. In
Section~\ref{sec:understanding} we discuss the origins for the various features
in the acquired spectra. The data in these first three subsections were
collected using the uncoated gold absorber. In Section~\ref{sec:resmater} we
present the effects of the MMC absorber material on the shapes of the detected
spectra. Then, in Section \ref{sec:resacet} we present exploratory observations
revealing neutral fragmentation products from the acetone radical cation, the
largest molecule studied here.

\subsection{MMC detection of atomic ions}\label{sec:resato}

\begin{figure*}
    \centering
 	\includegraphics[width=0.9\textwidth]{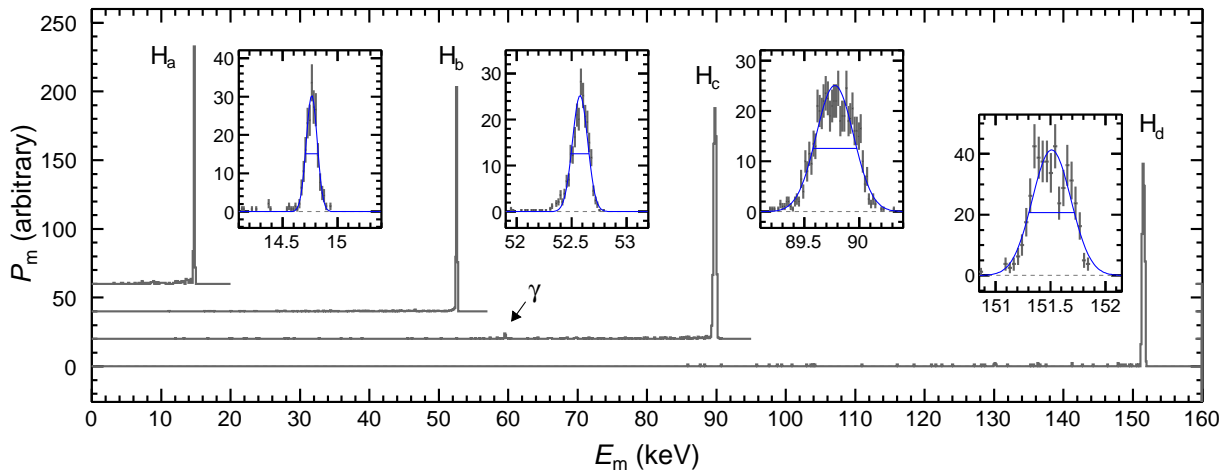}
	\caption{\label{fig:ProtEdistr}The MMC-detected energy spectra $\Pm$ for
proton
beams at $\Eib=14.7$, 52.6, 89.9, and 151.5~keV are plotted in
gray as a function of the detected energy $\Em$. We label the spectra as $\rm
H_a$, $\rm H_b$, $\rm H_c$, and $\rm H_d$, respectively. The values of $\Pm$ are
given on an arbitrary scale.   In order to visually separate the spectra,
vertical offsets have been added to $\rm H_a$, $\rm H_b$, and $\rm H_c$. The
$\gamma$-source signal is labeled in $\rm H_c$.  The insets display the spectral
details close to the respective beam energies $\Eib$. The statistical error bars
are displayed only in the insets. Gaussian-fit curves of the dominant peak in
each spectrum are plotted in blue. The blue horizontal solid lines indicate the
respective FWHMs and the dashed lines the baselines.}
\end{figure*}

In Figure~\ref{fig:ProtEdistr} we present four $\Pm(\Em,\Eib,{\rm H}^+)$ spectra
acquired with proton beams at $\Eib=14.7$, 52.6, 89.9, and
151.5~keV. In the figure we label the spectra as $\rm H_a$, $\rm H_b$, $\rm
H_c$, and $\rm H_d$, respectively. Each spectrum is dominated by a sharp peak
close to $\Eib$. A fit of the spectra in the vicinity of the peak using a
Gaussian function gives FWHMs of  $\Delta_{\rm m} = 117\pm7$, $165\pm29$,
$401\pm10$, and $415\pm26$~eV. The errors given are the statistical
uncertainties only.

Each of the acquired spectra is a convolution of the detector energy-response
function and the ion energy distribution:
\begin{equation}
 \Pm(\Em,\Eib,M) = \Fdet(\Em,\Ei,M)\otimes\Pib(\Ei,\Eib). \label{eq:PmConv}
\end{equation}
For the $\rm H_a$ and $\rm H_b$ spectra, voltage source A has been employed to
accelerate the proton beams. In this case, the ion energy distribution width
$\Delta_{\rm ib}^{\rm A}$, which has been determined independently (see
Section~\ref{sec:ionprod}), is negligibly small compared to the intrinsic
detector resolution $\Delta_{\rm det}^{\rm 0}$ derived in Section
\ref{sec:DetRespFnc}. Hence, the ion beam energy distribution $\Pib$ can be
neglected in Equation~(\ref{eq:PmConv}). Correspondingly, the $\rm H_a$ and $\rm
H_b$ spectra reflect directly the overall energy-response function $\Fdet$ and
the corresponding overall detector resolution is given by  $\Delta_{\rm
det}=\Delta_{\rm m}$. On the other hand for the spectra $\rm H_c$ and $\rm H_d$,
acquired with voltage source B, the widths of the prominent peaks are larger by
a sizable factor, in spite of the only slightly higher ion energies. We conclude
that the widths of these spectra are strongly influenced by the ion beam energy
spread. From separate measurements (explained in Section~\ref{sec:ionprod}) we
derived a value  of $\Delta{\rm_{ib}^B}=320\pm100$ eV for this energy spread.
From the measured widths, and approximating both the measured spectra $\Pm(\Em)$
and the ion energy distribution $P_{\rm ib}(\Ei)$ as normal distributions, we
obtain the detector resolution as
\begin{equation}
 \Delta_{\rm det} \approx \sqrt{(\Delta_{\rm m})^2 - (\Delta_{\rm ib}^{\rm
B})^2.}
\label{eq:FWHMsubtr}
\end{equation}
For the $\rm H_c$ and $\rm H_d$ spectra, we find $\Delta_{\rm det} =
241\pm17$~eV and $264\pm41$~eV, respectively. The given errors reflect only the
statistical uncertainties of the $\Delta_{\rm m}$ widths. The inferred width
$\Delta_{\rm det}$ resulting from Equation~(\ref{eq:FWHMsubtr}) must be larger
than the intrinsic detector resolution $\Delta{\rm_{det}^0}=107$~eV. This
requirement and the measured $\Delta_{\rm m}$ for spectra  $\rm H_c$ and  $\rm
H_d$ set an additional limit on the ion energy spread to be $\Delta_{\rm
ib}^{\rm B}<386$~eV. This enables us to reduce the estimated systematic
uncertainty on the beam energy spread $\Delta_{\rm ib}^{\rm B}$ to
$^{+66}_{-100}$~eV.

\begin{figure}
    \centering
 	\includegraphics[width=0.9\columnwidth]{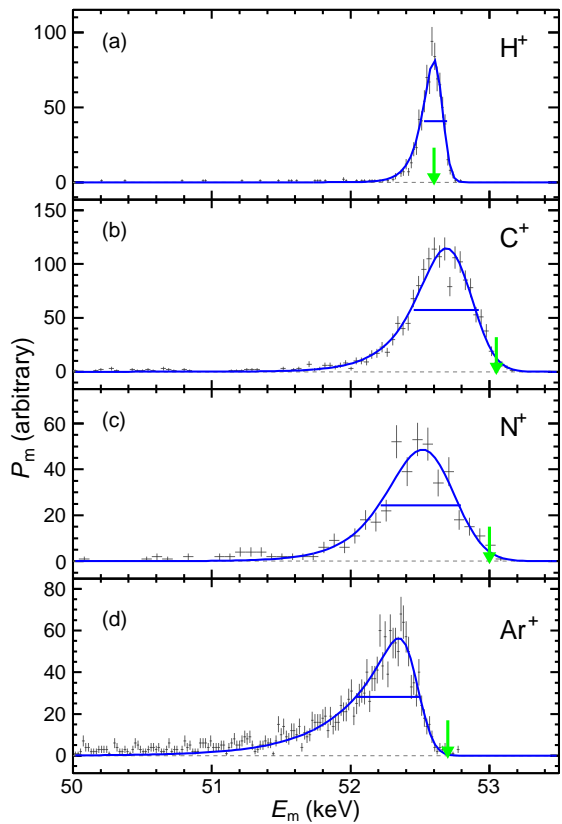}
	\caption{\label{fig:AtEdistr}The MMC-detected energy spectra $\Pm$ for
H$^+$, C$^+$, 
N$^+$, and Ar$^+$ beams at beam energies around $\Eib=53$~keV are shown by gray
data points in panels (a) to (d), respectively. The data for H$^+$ and Ar$^+$
were acquired using voltage source A, while data for C$^+$ and N$^+$ were
acquired using voltage source B. The statistical uncertainties are given by
the vertical error bars. Green arrows mark the respective nominal ion beam
energies $\Eib$. Fits by Equation~(\ref{eq:fitasym}) are shown by the blue
curves. The blue horizontal lines indicate the respective FWHMs.}
	
\end{figure}

In addition to protons, we have studied the MMC resolution for the atomic ions
C$^+$, N$^+$, and Ar$^+$. In Figure~\ref{fig:AtEdistr} we compare the
corresponding energy spectra $\Pm(\Em)$ acquired at  beam energies around
$\Eib=53$~keV. Similar to the proton spectra, the spectra of the heavier atomic
ions are also dominated by a peak close to $\Eib$. In contrast, however, this
main peak now has a clear tail extending towards low energies, suggesting that a
 small but significant part of the ion kinetic energy was not detected for a
noticeable fraction of events with heavier atomic ions. In order to
quantitatively describe the resulting asymmetric peak shape we fit the spectra
by a convolution of a Gaussian function and a left-sided exponential function
\cite{Horansky:JAP:2010,Bortels:IJRAI:1987}
\begin{eqnarray}
 f(\Em) = && \frac{A}{2\tau} \exp \left(
\frac{\Em-E_0}{\tau}+\frac{\sigma^2}{2\tau^2}\right)\nonumber\\
&& \times {\rm erfc}\left[ \frac{1}{\sqrt{2}}\left(
\frac{\Em-E_0}{\sigma}+\frac{\sigma}{\tau}\right)\right],
\label{eq:fitasym}
\end{eqnarray}
where $A$ is the peak area, $\tau$ is the exponential decay parameter of the
left-sided asymmetric part, $E_0$ is the peak position, $\sigma$ is the
Gaussian width, and $\rm erfc$ is the complementary error function. The
fitting range was iteratively adjusted to $E_0-3\sigma-3\tau \leq \Em \leq
E_0+3\sigma$. 

\begin{figure*}
    \centering
 	\includegraphics[width=0.41\textwidth]{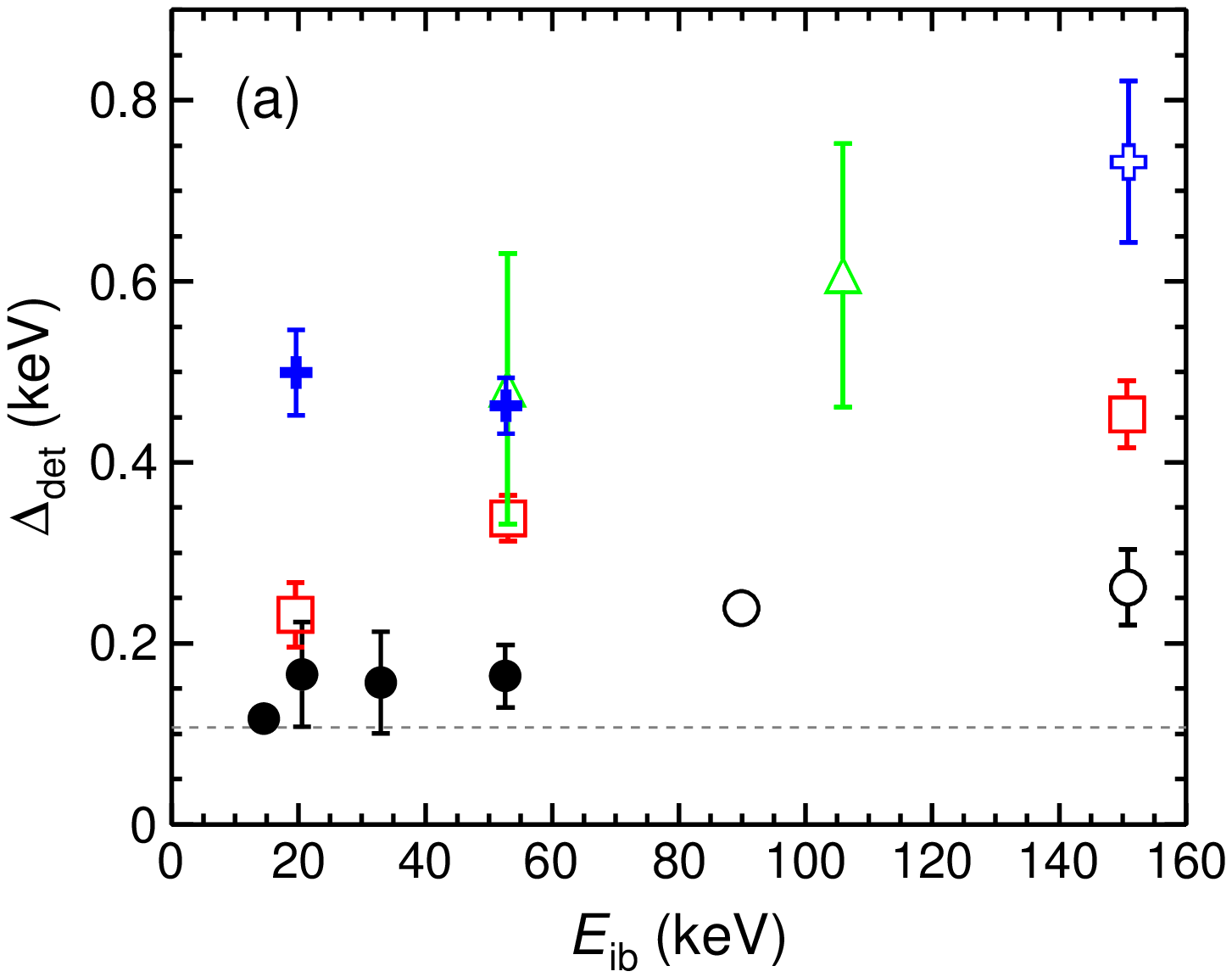}
 	\includegraphics[width=0.473\textwidth]{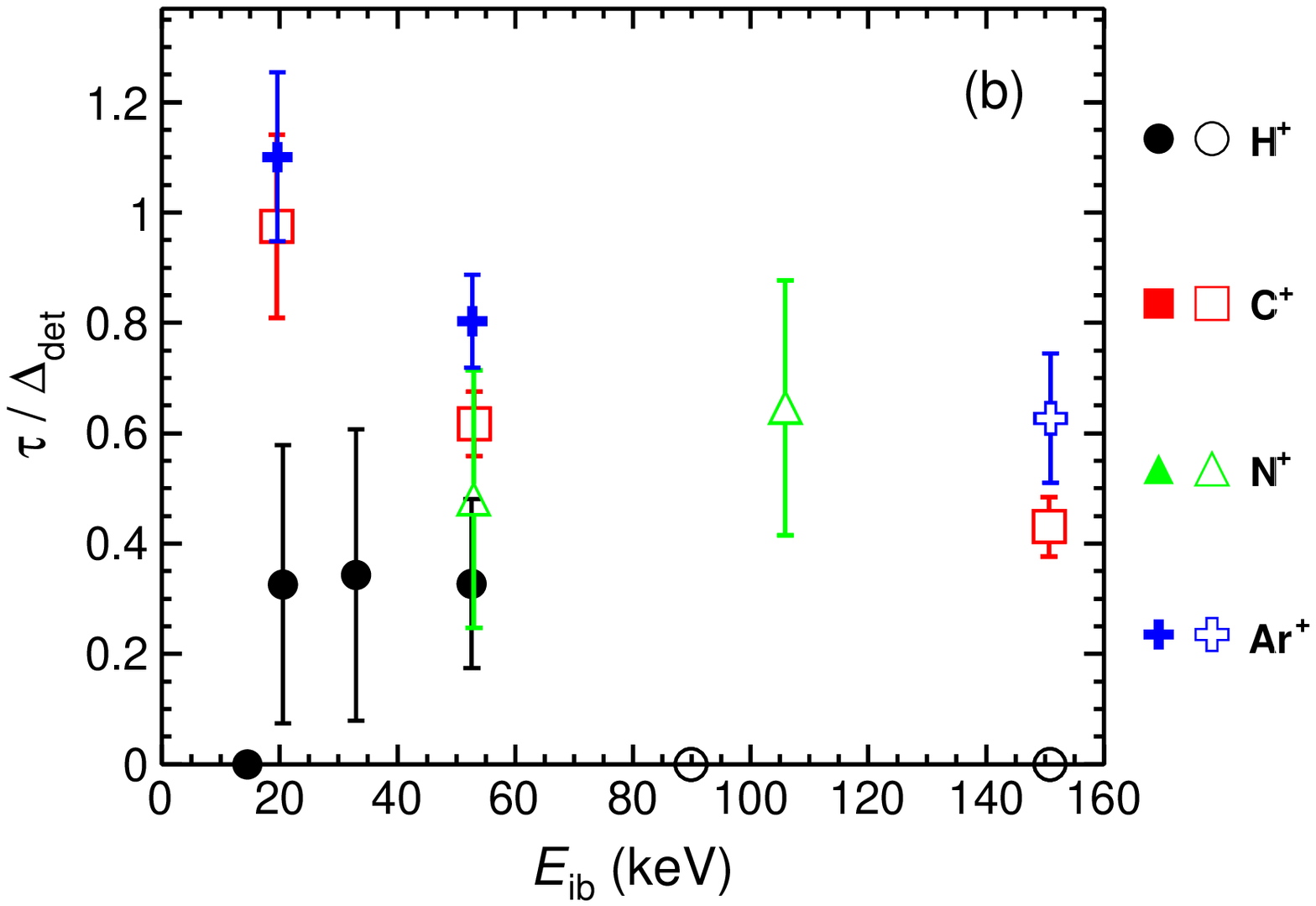}
	\caption{Panel (a):
 The widths $\Delta_{\rm det}$ of the dominant peak feature in the MMC-detected
energy spectra for various atomic ions are displayed as a function of the ion
beam energy $\Eib$. The legend for ion types is displayed on the very right-hand
side. The ion beams for the full-symbol data were accelerated by voltage
source~A. The correspondingly low ion energy spread was neglected. The hollow
symbol data were acquired with a less stable beams using voltage source~B. These
widths were corrected for the ion energy spread using
Equation~(\ref{eq:FWHMsubtr}). The error bars reflect the statistical
uncertainties of the fitted widths, but do not include the $^{+66}_{-100}$~eV
systematic error on the ion energy spread. The dashed line indicates the average
intrinsic detector resolution of $\Delta^0_{\rm det}=107$~eV. Panel (b): The
peak asymmetry expressed as the ratio of the left-sided exponential decay
parameter and the peak width ($\tau/\Delta_{\rm det}$) is plotted as a function
of $\Eib$. The meaning of the full/hollow symbols and of the error bars is
identical to panel~(a). The zero amplitude of some points (protons only)
indicate fully symmetric peaks where the fit by Equation~(\ref{eq:fitasym})
failed for numerical reasons.
}
	\label{fig:AtFWHMs}
\end{figure*}

In Figure~\ref{fig:AtFWHMs} we present the peak widths $\Delta_{\rm det}$ and
the asymmetry parameters $\tau/\Delta_{\rm det}$ for all of the atomic
ion spectra. The measured  width of each spectrum, $\Delta_{\rm m}$, is the
numerically determined FWHM of the fitted function $f(\Em)$. As explained above
for protons, the detector resolution for measurements with
voltage source A  is taken as  $\Delta_{\rm det}=\Delta_{\rm m}$. Similarly, we
have corrected the  widths $\Delta_{\rm m}$ for spectra acquired with the
voltage source B using the approximate method given by
Equation~(\ref{eq:FWHMsubtr}).

The derived detector resolutions $\Delta_{\rm det}$ for atomic ions, Figure
\ref{fig:AtFWHMs}(a), display a clear dependency on the ion type.  The quantity
$\Delta_{\rm det}$ grows with increasing ion mass. The best MMC energy
resolution was achieved for proton beams with $\Delta_{\rm det}<200$~eV at all
measured values of $\Eib$. The narrowest spectrum was found with protons
at $\Eib=14.7$~keV,  giving $\Delta_{\rm det}=117\pm7$~eV. The best relative
resolution was achieved with protons at $\Eib= 151.5$~keV, namely $\Delta_{\rm
det}/\Eib\approx1.7\ee{-3}$. The peak widths increase for heavier atoms,
reaching $\Delta_{\rm det}\approx 700$~eV for Ar$^+$. Given the large
statistical uncertainties and the systematic uncertainty in $\Delta{\rm_{ib}^B}$
(relevant for the hollow symbols), no clear dependency of $\Delta_{\rm det}$ on
$\Eib$ can be identified.

In Figure~\ref{fig:AtFWHMs}(b) we show the peak asymmetry described  by the
ratio of the left-sided exponential decay constant $\tau$ to the peak FWHM
(i.e., $\Delta_{\rm det}$). The most symmetric spectra - nearly Gaussian - were
acquired with protons. Fitting these spectra using Equation~(\ref{eq:fitasym})
failed in some cases due to numerical divergence at $\tau\rightarrow0$. In such
cases we have used a pure Gaussian to fit the data and report a generic value
for $\tau$ of ``0''. Heavier atoms, compared to the protons, display more
asymmetric peaks with a slightly decreasing tendency for increasing $\Eib$. 

\begin{figure}
    \centering
 	\includegraphics[width=\columnwidth]{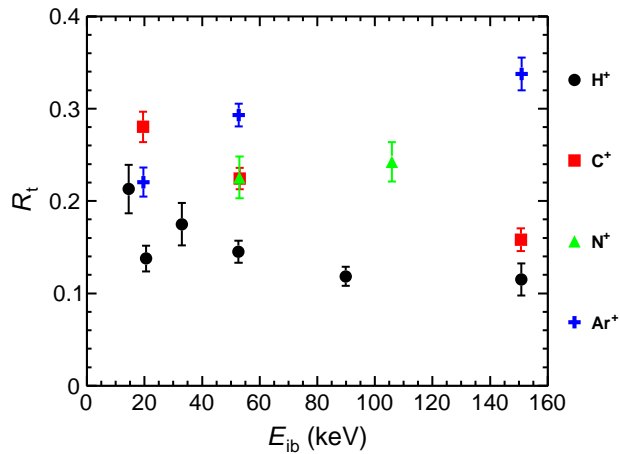}
	\caption{
The fraction of events detected at energies below
the dominant peak, $R_{\rm t}$, is plotted as a function of ion beam energy
$\Eib$ for various atomic ion beams. The exact definition of the $R_{\rm t}$
fraction is given in the text.  The error bars indicate statistical
uncertainties.}
	\label{fig:AtBacksc}
	
\end{figure}

In all of the atomic ion measurements, most events were detected within the
main peak, i.e., at energies of $E_0-3\sigma-3\tau \leq \Em \leq E_0+3\sigma$.
At higher energies no counts were recorded, which demonstrates the
extremely low-background detection capability of MMC detectors. In the
low-energy tails, the spectral count rates are small, decreasing smoothly
towards lower energies (ignoring the events at $\sim 59.5$~keV from the $\gamma$
source). In Figure~\ref{fig:AtBacksc} we plot the fractions of
event counts detected in these far low-energy tails, defined as
\begin{equation}
 R_{\rm t} =
\frac{\int_0^{E_0-3\tau-3\sigma}\Pm(\Em)\,d\Em}{\int_0^{\infty}\Pm(\Em)\,d\Em}.
\end{equation}
$R_{\rm t}$ reaches values as low as $\sim10\%$ for  high-energy proton
projectiles. This fraction grows to up to $\sim 35\%$ for heavier atoms. For
H$^+$ and C$^+$ a clear decrease of $R_{\rm t}$ with increasing $\Eib$ can be
seen. This trend cannot be seen in the N$^+$ data and even seems to be inverted
for Ar$^+$ ions.  Measurements presented in Section~\ref{sec:resmater} show
that the tail fraction is significantly reduced with the Al-coated absorber.

\subsection{MMC mass spectrometry on molecular ions}\label{sec:masspec}
\begin{figure*}
    \centering
 	\includegraphics[width=0.9\textwidth]{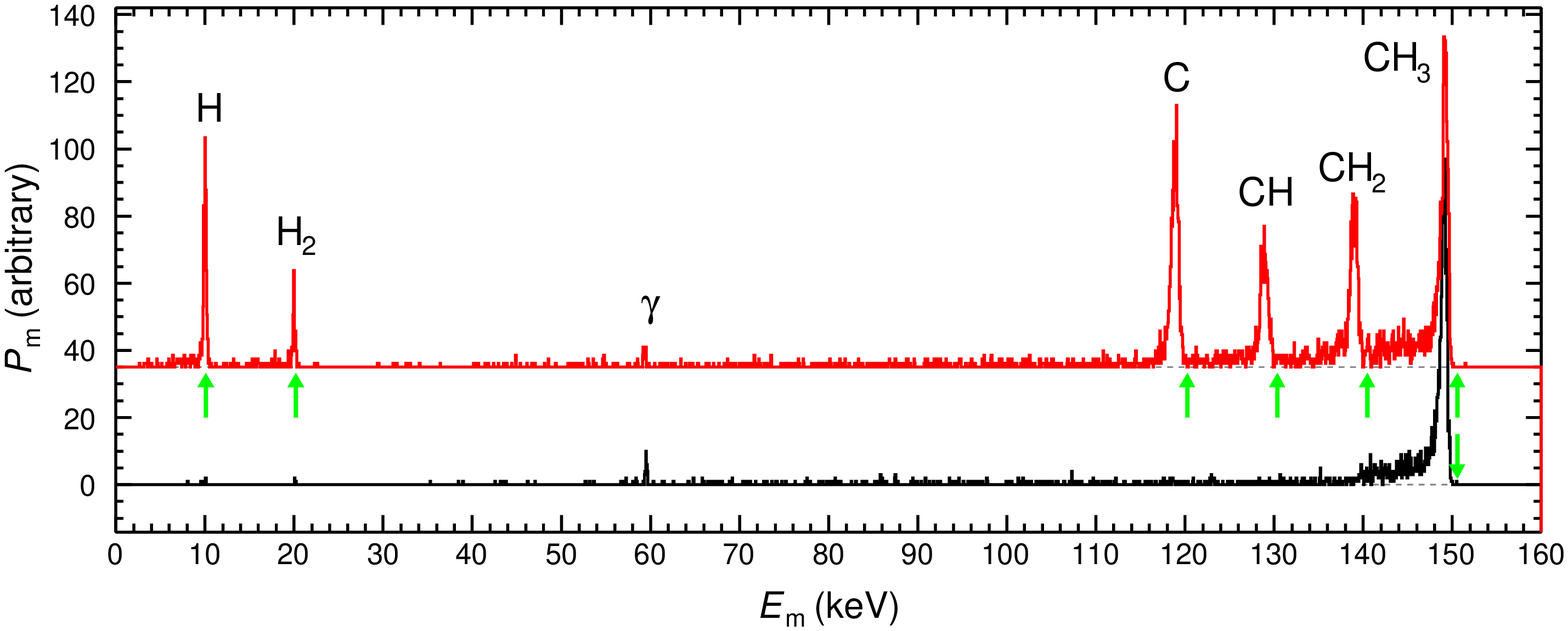}
	\caption{The MMC-detected energy spectra $\Pm$ for a CH$_3^+$ 
beam at $\Eib=150.6$~keV is plotted by the black line as a function of the
detected energy $\Em$. The red curve plots the spectrum acquired with ions
prevented from reaching the detector, so that only neutral CH$_3$ and its
neutral fragments were detected. The data have been vertically shifted from zero
in order to visually separate the two spectra. The rightmost green arrow
indicates the nominal ion beam energy $\Eib$  and the remaining green arrows,
going from left to right, show the energy fraction corresponding to the masses
of H, H$_2$, C, CH, and CH$_2$ fragments.
}
	\label{fig:CH3neutr}
\end{figure*}

In addition to atomic ions we have also studied the calorimetric detection for
several types of molecular ions. As an example we plot the detected energy
spectrum for a beam of CH$_3^+$ at $\Eib=150.6$~keV in
Figure~\ref{fig:CH3neutr} (black line). Similar what we found for atoms, this
molecular spectrum is dominated by a single peak near to $\Eib$. From the peak
width of $\Delta_{\rm m}=786\pm71$~eV and from the
ion beam energy spread $\Delta_{\rm ib}^{\rm B}$, we derive a FWHM
detector resolution of $\Delta_{\rm det} = 718\pm78$~eV. This exceeds the energy
width for C$^+$ atomic ions (Figure~\ref{fig:AtFWHMs}) by $\sim200$~eV.

Using an energetic molecular ion beam allowed us to demonstrate the MMC mass
spectrometry capabilities directly in a single measurement. For this we directed
all of the charged particles away from the detector, using a
magnetic deflector placed $\sim1$~m upstream of the detector. In this
configuration, the only
remaining particles hitting the detector were neutral atoms and molecules
resulting from collisions of the molecular ions with residual gas in the ion
beam line. Clearly, only a small fraction
of the collision-induced neutrals are able to reach the MMC detector through the
narrow pinhole in the mask in front of the detector. Thus, to reach acceptable
data quality from this measurement configuration we have used high ion beam
intensities and increased residual gas pressure in the beamline.
Processes leading to neutral fragments are discussed in
Section~\ref{sec:resacet}.

The red curve in Figure~\ref{fig:CH3neutr} displays the energy spectrum of
neutral fragments resulting from collisions of CH$_3^+$ ions at $\Eib=150.6$~keV
with residual gas molecules along the upstream beam line and reaching the
detector pixel through the pinhole mask. The spectrum displays distinct peaks at
energies matching well to the respective fragment-to-parent mass fractions of
$\Eib$ for all possible fragments of the CH$_3^+$. The peak widths in the
neutral spectrum are $\Delta_{\rm m}=333\pm20$~eV for H, $349\pm46$~eV for
H$_2$, $908\pm180$~eV for C, $1230\pm67$~eV for CH, $1013\pm155$~eV for CH$_2$,
and $813\pm78$~eV for CH$_3$. The CH$_3$ peak width is equal to that from the
CH$_3^+$ spectrum (to within the statistical uncertainty). This suggests that
the molecular velocity is not significantly altered in the non-dissociative
neutralizing collision with the residual gas. Moreover, this also allows us to
conclude that the charge of the detected particle is irrelevant for the
calorimetric measurement, to within the detector resolution. 

Interestingly, the peaks for the CH and CH$_2$ fragments are clearly broader
than that for CH$_3$. The additional spread may originate from the kinetic
energy released in the CH$_3$ dissociation process. For example, for a 150~keV
CH$_3$ beam, a center-of-mass kinetic energy of 1~eV released into CH and H$_2$
fragments dissociating along the ion beam axis transforms into an $\sim260$~eV
change of the fragment kinetic energy in the laboratory frame. Since the present
measurements detect only a small fraction of the neutral fragments produced in
the beam, the relative detection yields for the different fragment species may
depend significantly  on the energy released and the angular emission pattern of
the respective fragmentation channels. Nevertheless, we find that even under
these non-ideal conditions the high-resolution calorimetric mass spectra still
yield qualitative information about the molecular composition and the
fragmentation channels.

In a more pronounced fashion than for the atoms, the mass peaks for both the
primary molecular ion and the neutral molecular fragments are
accompanied by tails that will be further discussed below. These
wider tails turn out to be significantly influenced by the
coating of the absorber in which the molecular stopping occurs (see
Section~\ref{sec:resmater}).

\subsection{Modeling MMC energy spectra}\label{sec:understanding}
\begin{figure*}
    \centering
 	\includegraphics[width=0.9\textwidth]{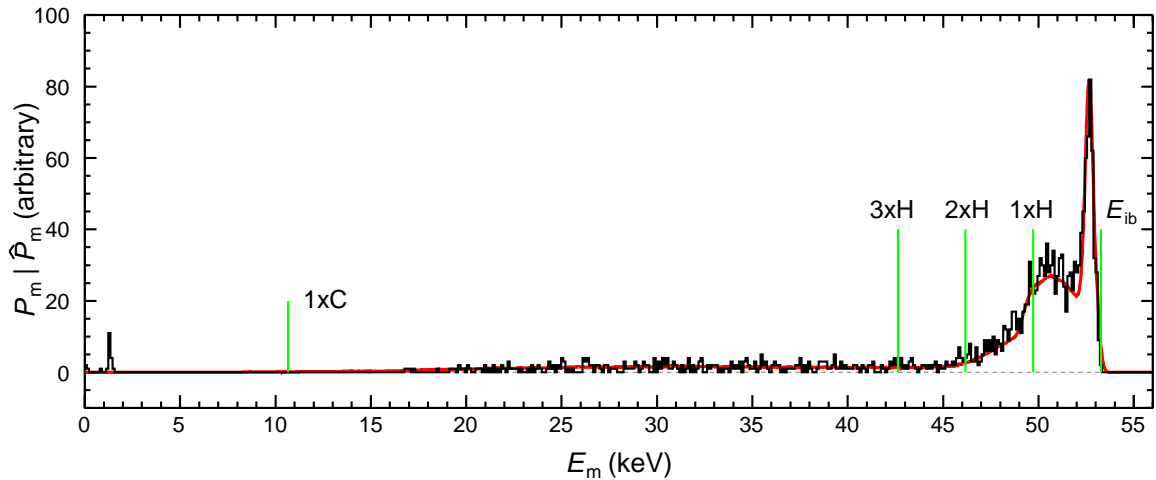}
	\caption{The MMC-detected energy spectra $P_{\rm m}$ for CH$_3^+$ beam
at $\Eib= 53.3$~keV is plotted by the black line as a function of detected
energy $\Em$. The red curve represents our SRIM-based model spectrum
$\hat{P}_{\rm m}$, assuming a Frenkel-pair energy $E_{\rm FP}=2.8$~eV. Going
from right to left, the green vertical lines mark the nominal ion beam energy
$\Eib$  and the maximal energy losses due to backscattering of one, two, and
three H-atoms, and one C-atom. The model spectrum is scaled in amplitude so that
the integral between $\Em=5$~keV and 54~keV matches the experimental data.
}
	\label{fig:CH3model}
	
\end{figure*}

As was shown above (Section~\ref{sec:resato}), the detected energy
spectra for atomic projectiles are dominated by a single narrow peak close to
the nominal ion beam energy $\Eib$, while more complex shapes are found for 
molecular
projectiles. Thus, the CH$_3^+$ data at $\Eib=150.6$~keV (Figure
\ref{fig:CH3neutr}, black line) display a low-energy shoulder on the main peak
in the energy range from $\sim 140$ to 147~keV. We find that the
relative intensity of this feature grows significantly for lower ion beam
energies, as seen in the CH$_3^+$ data acquired at
$\Eib= 53.3$~keV (Figure~\ref{fig:CH3model}). Similarly, we also observed
prominent low energy shoulders in energy spectra  recorded for 
CO$_2^+$ and (CH$_3$)$_2$CO$^+$ beams. 

In order to understand the detected spectral shapes we have modeled the
processes occurring during the stopping of molecular projectiles in the MMC
absorber. In our simplified  model we assume that the detected molecule breaks
into
separate atoms immediately after impinging on the detector surface and neglect
the few eV of molecular binding energy. Additionally, we disregard the charge
state of the projectile particles, based on our findings described earlier. With
these assumptions we can study stopping processes in the absorber separately for
each of the atomic constituents of a given molecule, such as  C and H in case of
CH$_3^+$. The model for the molecular spectra can then be constructed from a
suitable combination of the modeled atomic spectra.

We have employed the SRIM software to simulate the energy loss processes of the
projectile atoms in  the MMC absorber. SRIM \cite{Ziegler:NIMB:2010} is a 3D
Monte-Carlo code that simulates the propagation of ions in matter. It describes
not only collisions between the projectile and target atoms, but also accounts
for target-target collisions due to cascades triggered by recoiling target
atoms, backscattering of projectiles,
and sputtering  of atoms from the target material. The resulting detailed data
on each collision allows us to evaluate the amount of projectile kinetic energy
which is not transformed into heat and thus escapes calorimetric detection.
Specifically, for each simulated detection event we have calculated whether the
projectile atom was back-scattered, and if yes, then what fraction of the
kinetic energy has remained in the absorber. Next, we have evaluated the number
and the total kinetic energy of target atoms sputtered from the detector
surface. Lastly, for each impact we track the number of defects created in the
absorber lattice. Such defects arise when an atom of the target lattice is
kicked out by recoil from its stable position (either by the projectile atom or
another recoiled target atom). The resulting pair of an interstitial atom and a
vacancy in the lattice (i.e., a Frenkel pair or a Frenkel defect
\cite{Kittel:book:2005}) holds a potential energy $E_{\rm FP}$. This non-thermal
energy storage results in a reduction of the energy detectable by the MMC. The
SRIM code calculates the number of Frenkel defects along each incident ion
track. To obtain the total energy loss due to Frenkel pair formation we multiply
the number of Frenkel pairs by $E_{\rm FP}$ which is a parameter of our model.
The number of replacements where an energetic target atom terminates its
movement in a previously created vacancy, is negligible. Altogether, for each
impact event we have subtracted the above listed energy losses from the initial
kinetic energy of the incident atom and thus obtain individual atomic detected
energies $\Em^{\rm ato}$. For a given molecule, we simulate $10^4$ impact events
related to each atomic constituent type. This yields the
model energy spectra for each atomic species.

An additional loss of energy may arise from secondary electron emission by
the detector absorber after the impact of a particle. Studies on
secondary electron emission from solid gold targets, however, have found only
small secondary electron yields of $\lesssim 2$ electrons per incident ion
\cite{Veje:NIMPR:1982,Lorincik:PRB:2000} and mean secondary electron energies
of $\sim 10$~eV \cite{Hasselkamp:NIMB:1986} for a broad range of
projectile energies and projectile types. Based on those studies we estimate the
energy loss due to secondary electron emission to be negligible compared to
the other energy loss processes and we do not include it in our model.

To calculate the detectable energy for a molecular projectile, $\Em^{\rm
mol}$, we have randomly picked simulated events for each of the constituent
atoms and sum the corresponding $\Em^{\rm ato}$ values. Then we have added a
random Gaussian spread
describing the intrinsic detector resolution ($\Delta_{\rm det}^{0}=107$~eV) and
the ion energy spread ($\Delta_{\rm ib}^{\rm B}=320$~eV FWHM), obtaining the
model energy distribution $\hat{P}_{\rm m}(\Em)$. Adding the ion beam energy
spread only in the last step (instead of varying $\Ei$ as the input parameter of
the SRIM simulations) is justified by the small relative ion beam energy spread
of $\Delta_{\rm ib}^{\rm B}/\Eib < 1\%$, within which the relative contributions
of the various energy loss processes stay nearly constant. 

\begin{figure*}
    \centering
 	\includegraphics[width=0.9\textwidth]{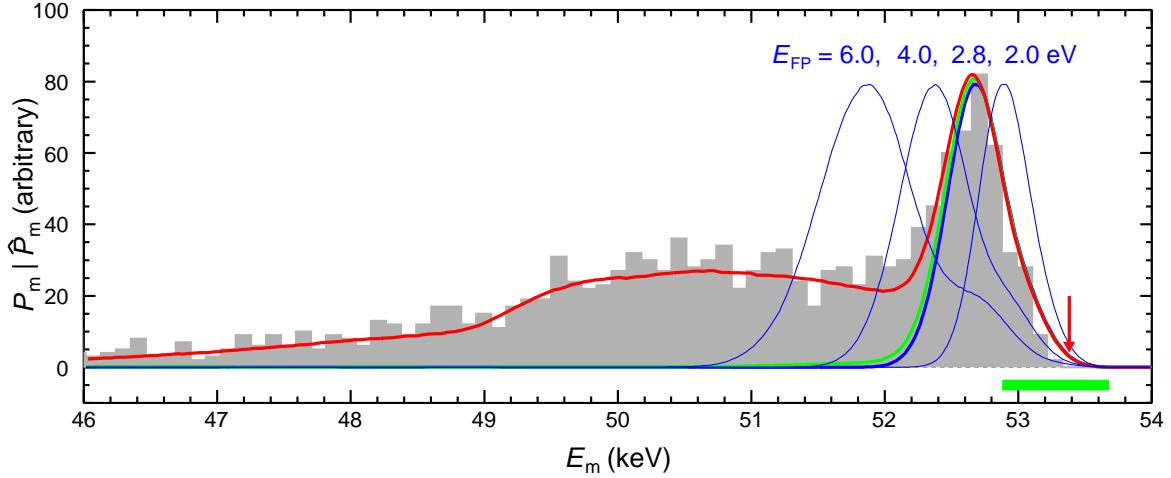}
	\caption{A comparison of the MMC detected energy spectrum (gray) and a
corresponding SRIM-based model for CH$_3^+$ (red). The thick blue line plots the
model spectrum assuming energy losses only via Frenkel-pair formation with an
adjusted Frenkel-pair energy of $E_{\rm FP}=2.8$~eV. The thin blue lines plot
the same model using an alternative value for $E_{\rm FP}$ of 6.0~eV, 4.0~eV,
and 2.0~eV, respectively. The green line plots the model assuming Frenkel-pair
formation and sputtering. The difference between red and green line then
reflects the effect of backscattering. The horizontal green bar labels the
uncertainty range for the nominal ion beam energy $\Eib=53.3\pm0.4$~keV in the
experiment. The red arrow marks the adjusted ion beam energy for the model.}
	\label{fig:CH3modelZoom}
	
\end{figure*}

In Figures~\ref{fig:CH3model} and \ref{fig:CH3modelZoom} we compare the model
energy distribution $\hat{P}_{\rm m}$ for CH$_3^+$ at $\Eib=53.3$~keV to
the corresponding  experimental data. In red we plot the result of the complete
model including all of the energy loss processes mentioned above and for the 
value of $E_{\rm FP}$ yielding the best agreement with the data. Partial effects
are shown in Figure~\ref{fig:CH3modelZoom} for the Frenkel-pairs only (blue) as
well as for Frenkel pairs plus sputtering (green). Comparing this latter curve
with the full result (red) shows the rather strong effect of the backscattering
process. The curves for the Frenkel-defects only are also shown for a range of 
$E_{\rm FP}$ values.

The shape of the dominant peak at $\Em=52.7\pm0.5$~keV can be assigned to the
energy spread due to Frenkel pair formation. The Frenkel-pair energy used in the
model affects only this feature in its width and mean energy. Comparing directly
the peak energy in the model to the measured data is difficult due to the
systematic uncertainty in the nominal ion beam energy $\Eib$
(Sec.~\ref{sec:ionprod}). Instead we have varied the $E_{\rm FP}$ value in steps
of 0.05~eV and in each iteration compared the width of the dominant peak in the
model spectra to that in the measured data. Simultaneously we adjusted the model
beam energy so that the center of the peaks overlap. The best match was reached
for $E_{\rm FP} = 2.8$~eV and by adjusting the model beam energy by $+0.1$~kV
(this is only 0.2\% of $\Eib$) with respect to the nominal beam energy of
$\Eib=53.3$~keV. Such an energy shift is well below the estimated $\Eib$
uncertainty. The energy losses due to sputtering of absorber atoms add only a
small left-handed tail on the main peak and do not produce any distinct
spectral feature. Backscattering of H atoms is  responsible for the wider
feature between $\sim 45$ and 52~keV and backscattering of C for the tail
remaining visible down to $\sim 15$~keV. The small feature at $\sim 1.2$~keV was
not observed in most of other acquired spectra and is interpreted as a
spurious event not originating from CH$_3^+$. 

Backscattering appears to be the key for explaining the differences between the
atomic and molecular spectral shapes beyond the main peak. For atomic
projectiles, the  observed low-amplitude and low-energy tails are attributed to
backscattering of the projectile atoms themselves; any of the back-scattered
atoms can take away up to the full original kinetic energy. Typical spectra can
be seen in Figure~\ref{fig:ProtEdistr}. On the other hand, for molecular
projectiles the dissociated atomic fragments are expected to backscatter
independently, each fragment being able to carry away only the part of the
original projectile kinetic energy given by the fragment-to-molecule mass ratio.
In the case of CH$_3^+$, energy loss maxima of $\Eib/15$ and of $12\Eib/15$
result for backscattered single H and C atoms, respectively. Given the H
multiplicity, the most likely backscattering process is that of a single H
fragment, taking away energy between 0 and $\sim\Eib/15$. As a consequence we
observe a strong, broad feature in $\Pm$ between $\Em\approx \Eib-\Eib/15$ and
$\Em\approx\Eib$. Structures in  $\Pm$ at lower energies decrease in magnitude
as the simultaneous backscattering of multiple H atoms becomes less probable.
The lowest energies can be accessed only by backscattering of carbon atoms.

Only a few parameters enter our model. The most influential one is the
Frenkel-pair energy which directly scales the width and position of the main
peak. Our fitted value $E_{\rm FP}=2.8$~eV  has a strong correlation with the
beam energy spread $\Delta_{\rm ib}^{\rm B}$. The $^{+66}_{-100}$~eV uncertainty
of $\Delta_{\rm ib}^{\rm B}$ causes an uncertainty of $^{+0.9}_{-0.6}$~eV in the
adjusted value of $E_{\rm FP}$. 

Another important parameter entering the SRIM model is the threshold
displacement energy $E_{\rm displ}$, i.e., the minimum energy needed for a
recoiling target atom to overcome the lattice barrier, move away from its
original position, and form an interstitial. In our model we use $E_{\rm
displ}=43$~eV, a value recommended in a recent review \cite{Broeders:JNM:2004}.
The $E_{\rm displ}$ values listed in that review range
from 30~eV to 44~eV. Using this range as the uncertainty in $E_{\rm displ}$
results in an uncertainty of $^{+0.1}_{-0.7}$~eV for $E_{\rm FP}$. 

Furthermore we set the lattice binding energy to $E_{\rm latt}=3.81$~eV
\cite{Kittel:book:2005} and the surface binding energy for gold to $E_{\rm
surf}=3.83$~eV \cite{Hildenbrand:JPC:1962}. In the SRIM simulation $E_{\rm
latt}$ is the energy that every recoiling target atoms loses when it leaves  its
lattice  site. $E_{\rm surf}$ is relevant for calculating sputtering. The
estimated 0.5~eV uncertainties on $E_{\rm latt}$ and $E_{\rm surf}$ propagate to
a negligible uncertainty in $E_{\rm FP}$. For details on the relations between
$E_{\rm displ}$, $E_{\rm latt}$, and $E_{\rm FP}$ see, e.g., the SRIM
documentation \cite{Ziegler:NIMB:2010} or Figure~11 in \cite{Horansky:JAP:2010}.

Adding all the systematic uncertainties together in quadrature gives a
Frenkel-pair energy of $E_{\rm FP}=2.8\pm0.9$~eV. To the best of our knowledge,
no other experimental values of $E_{\rm FP}$ have been published for gold.
Calculated values range from $E_{\rm FP}=3.0$ to 4.1~eV, depending on the
calculational method as well as on the specific lattice defect type formed
\cite{Bauer:PLA:1982, Wales:PRB:1994}. The Frenkel-pair energy determined from
our model matches well with these calculated values.

\subsection{Optimizing MMC resolution - absorber materials}\label{sec:resmater}
Practical usage of the MMC detector for spectrometry requires a narrow 
energy-response function. We have shown that the dominant peak in all of the
collected energy spectra is narrow and  that relative resolutions down to $\sim
10^{-3}$ can be reached. Nevertheless, the low-energy tails of the
energy-response function can significantly complicate the analysis of energy
spectra, especially for molecular projectiles. 

Our modeling studies indicates that the low-energy tails in the measured spectra
originate predominantly from backscattering. Thus, in order to optimize the MMC
 energy-response function, we have searched for absorber materials with low
backscattering probabilities. SRIM models show a general trend of lower
backscattering for materials composed of lighter atoms (i.e., lower atomic
number $Z$). Backscattering is determined predominantly by the first few surface
monolayers. Hence only a thin absorber coating of a low-$Z$ material should be
sufficient to improve the MMC energy-response function. Moreover, SRIM
calculations indicate that fewer Frenkel pairs should be created in low-$Z$
materials. Thus, provided that the Frenkel-pair energy does not depend on the
absorber material or is lower for a specific low-$Z$ material, then  the
dominant peak should also
narrow.

\begin{figure}
    \centering
 	\includegraphics[width=\columnwidth]{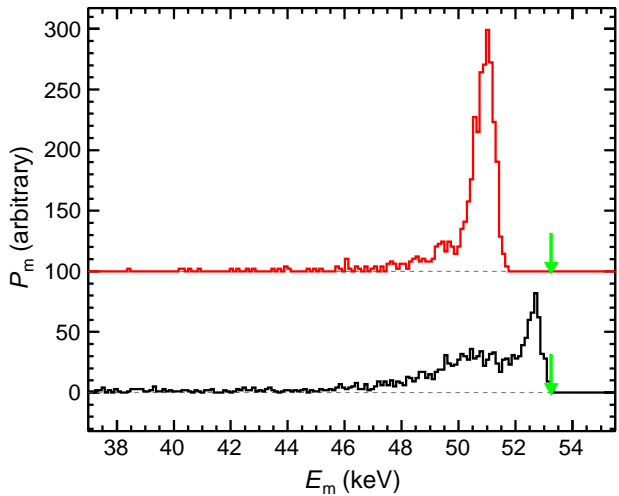}
	\caption{MMC detected energy spectra for CH$_3^+$ at $\Eib=
53.3$~keV, acquired with the gold-absorber pixel (black) and with the
aluminum-coated pixel (red). The amplitude of the spectra have been scaled such
that the total integrals are equal. An artificial  offset  in $\Pm$ has been
added for clarity. The green arrows label the nominal ion beam energy $\Eib$.}
	\label{fig:AuAlcomp}
\end{figure}

To test this hypothesis we have coated the 50~$\mu$m gold absorber of one of the
MMC pixels with 180~nm of aluminum. In Figure~\ref{fig:AuAlcomp} we compare the
energy spectra from the bare and Al-coated gold pixels, acquired with the same
beam of CH$_3^+$ at $\Eib=53.3$~keV. Clearly, using an Al coating results in a
strong suppression of the backscattering.  There is a clear reduction near 50
keV in the shoulder due to backscattering of H atoms, and also a clear decrease
below 46 keV in the tail due to backscattering of C atoms. Interestingly, the
main peak is shifted towards lower energies when using the Al-coated absorber.
This cannot be reproduced by our SRIM-based model. We anticipate that the shift
is due to the aluminum being superconductive at the MMC operating temperatures,
which is below the aluminum critical temperature of $T_{\rm c}=1.2$~K.  Hence,
the energy loss processes in the Al-coated pixel may be different from those for
normal-conducting gold. This interpretation is supported by previously published
experiments investigating the properties of various superconductors when used as
a particle or X-ray absorber in a micro-calorimeter. At first glance,
superconductors seem to be a very promising class of cryodetector materials,
considering their relatively small specific heat at temperatures well below
$T_\mathrm{c}$. However, as shown by~\cite{Cosulich:JLTP:1993}, which
summarized earlier work on various superconductors, and by more recent
studies~\cite{Porst:JLTP:2008,weldle:dipl:2008,porst:thesis:2010} on aluminum
and rhenium, the thermalization of energy in superconductors is not well
understood at temperatures as low as that of our absorber at $\sim 10$~mK. At
these temperatures there is very little phonon excitation ($T\lesssim
2\times10^{-4}\, \Theta_{\rm D}$, where $\Theta_{\rm D}$ is the Debye
temperature) and surprisingly long thermal pulse decay times have been
observed~\cite{Porst:JLTP:2008}. Analysis of these measurements suggests that a
significant fraction of the deposited energy, from a few percent to tens of
percent, could be stored in very long-lived excitations with lifetimes ranging
up to several seconds. This fraction of energy would escape detection. The
relevant long-lived excitations could be due to quasi-particles resulting from
broken Cooper pairs, excitations of magnetic flux lines, or excitations of
nuclear spins or atomic tunneling systems, to name just a few possibilities. In
fact, the apparent magnitude of the energy loss in our detector with the
aluminum coating, on the order of 5\%, fits well to these previous observations
and interpretations.

\begin{figure*}
    \centering
 	\includegraphics[width=0.9\textwidth]{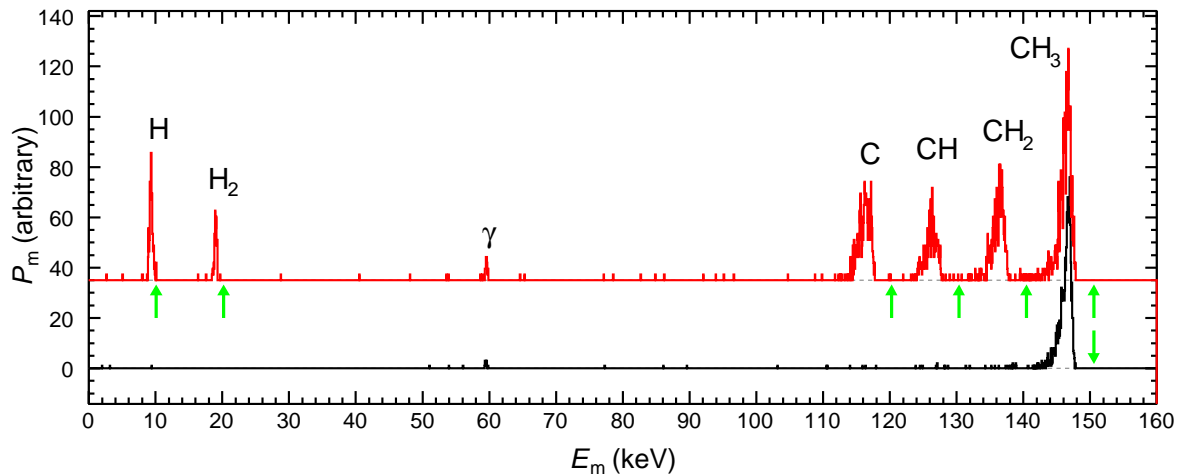}
	\caption{Same as Figure~\ref{fig:CH3neutr}, but acquired with
the Al-coated MMC absorber.}
	\label{fig:CH3neutrAl}
	
\end{figure*}

We also used the Al-coated detector pixel to reproduce the measurements on
neutral CH$_3^+$ fragments created by residual gas collisions of CH$_3^+$ ions
at $\Eib=150.6$~keV (see Section~\ref{sec:masspec}). The results are
displayed in Figure~\ref{fig:CH3neutrAl}. A comparison to the data acquired with
the bare gold absorber (Figure~\ref{fig:CH3neutr}) shows, again, a strong
reduction of the low-energy tail when the Al-coated absorber is used.

\subsection{Acetone radical fragmentation}\label{sec:resacet}
\begin{figure*}
    \centering
 	\includegraphics[width=0.9\textwidth]{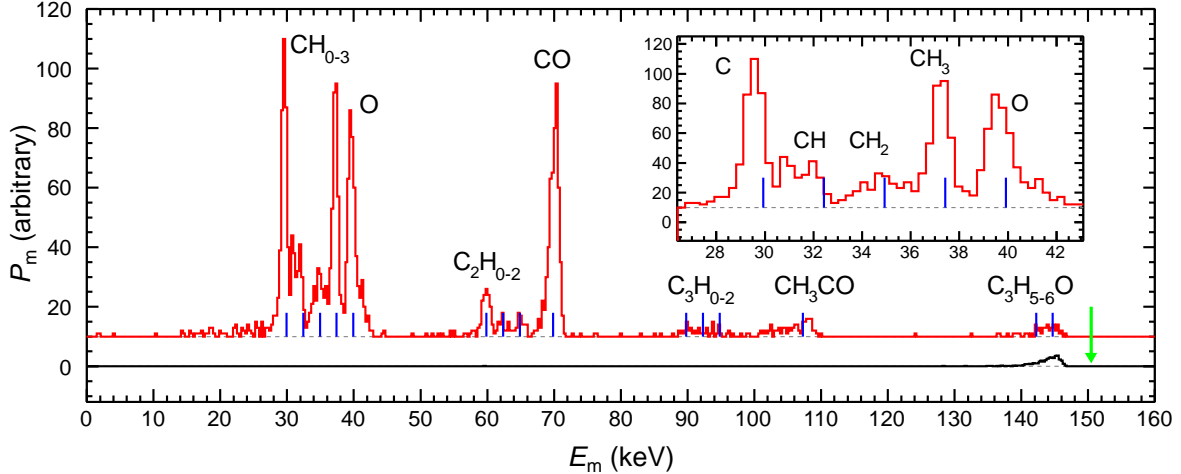}
	\caption{ Same as Figure~\ref{fig:CH3neutr}, but for acetone
radical ions (CH$_3$)$_2$CO$^+$ at $\Eib=150.45$~keV (black line) and for
neutral products from colliding these ions with residual gas (red curve). In
both cases the Al-coated MMC absorber was used. The blue vertical lines mark
the expected energies of the various neutral fragmentation products (see the
text for a detailed explanation). The green arrow indicates the nominal ion beam
energy $\Eib$. }
	\label{fig:AcetNeutrAl}
\end{figure*}

Using our MMC setup we have also investigated the fragmentation of acetone
radical cations, (CH$_3$)$_2$CO$^+$ (58~u), due to residual gas collisions.
To detect the neutral fragments we followed the approach outlined in
Section~\ref{sec:masspec}. The data acquired at $\Eib=150.45$~keV using the
Al-coated detector pixel are displayed in Figure~\ref{fig:AcetNeutrAl}. Similar
to our other measurements with this pixel, the measured energies are $\sim 4\%$
lower than those obtained with the gold-only pixel. For this large  molecule,
the data obtained with the uncoated gold pixel (not shown here) are much less
resolved than those from the Al-coated pixel, although, as was consistently
seen, the full-mass peak for the uncoated pixel lies much closer to $\Eib$. In
the neutral fragment spectrum of Figure~\ref{fig:AcetNeutrAl} (collected using
the Al-coated pixel) we assigned the most distinct peaks by multiplying the
fragment-to-molecule mass ratios by a full kinetic energy of only 96\% of $\Eib$
(the energy step corresponding to 1~u is then 2.49~keV). The small shifts
between the peak centers and the predicted fragment energies, seen for only a
few specific fragment masses (notably near 13~u at $\sim 32$~keV), may be due to
more complex, still unexplained effects in the stopping of the molecular
species.

Two processes are expected to generate the neutral fragments. The first is
collision induced dissociation (CID, also known as collision activated
dissociation - CAD), which does not change the charge of the target particle (a
residual gas molecule in our case). In CID at least one of the projectile
fragments stays charged. The second process is an electron transfer 
 from the neutral target to the
projectile cation neutralizing it. As in electron-ion
recombination, the ionization energy released then usually leads to the
dissociation of the neutralized projectile molecule into neutral fragments
(known as electron transfer dissociation - ETD). The electron transfer may also
result in
an internally excited, but non-dissociated, neutral molecule.

We also want to emphasize here that the setup used in these exploratory
studies was not designed for quantitative branching ratio measurements. The
kinetic energy released in the various CID and ETD outgoing channels may vary
significantly. This affects the transverse distribution of the
fragment impact positions on the detector. As a result, the true geometrical
efficiency for detecting the fragments through the pinhole in front of the
detector may vary strongly as well. Thus, the relative peak heights should  be
interpreted with care. Nevertheless, we find that the high mass resolution of
our setup provides unprecedented insight in the CID and ETD channels producing
neutral fragments by showing a number of discrete fragment mass peaks.

Many studies exist on CID and ETD of the acetone radical
\cite{Powis:IJMS:1979,Martinez:RCMS:1989,
Shukla:IJMS:1989,Martinez:JASMS:1992,Zhao:IJMS:1997}. However, the vast majority
of the techniques used collected only charged reaction products. Given this
limitation, these studies assigned the dominant dissociation channel to 
\begin{equation}
 {\rm(CH_3)_2CO^+} + A \rightarrow {\rm CH_3CO^+ + CH_3} + A. \label{eq:acet1}
\end{equation}
Here $A$ is the target particle, which in our case of a non-baked UHV beam line
is expected to be H$_2$O, N$_2$, CO, or H$_2$ \cite{ohanlon:book:2003}. We are
aware of only one study on acetone radical fragmentation capable of also
collecting neutral fragments. In specific, Ohkubo et al. \cite{Ohkubo:IJMS:2011}
used a cryogenic STJ detector in a novel mass spectrometry approach for their
investigations of 3~keV acetone radicals colliding with Xe atoms. They explained
their data by two additional ETD channels producing solely neutral fragments:
\begin{eqnarray}
 {\rm(CH_3)_2CO^+} + A \rightarrow {\rm CH_3CO + CH_3} + A^+,\\
 {\rm(CH_3)_2CO^+} + A \rightarrow {\rm CH_3 + CH_3 + CO} + A^+.\label{eq:acet3}
\end{eqnarray}
However, their relative resolution of only $\Delta E/E\approx0.14$ was not
sufficient to resolve molecular fragments differing only by the number of
attached hydrogen atoms.

Our results clearly show neutral fragments which do not originate from any of
channels $(\ref{eq:acet1})-(\ref{eq:acet3})$. The most interesting aspect of our
data indicates that there is significant production of single C and O atoms as
well as of the non-hydrogenated carbon molecules C$_2$ and C$_3$. In the
measurements of Ohkubo et al., due to their lower relative energy resolution,
any potential contribution from these fragments would be hidden in the peaks
assigned as CH$_3$, CO, or CH$_3$CO. A full comparison between their work and
ours certainly needs to take into account the differences between the two
studies including collision energy, target gas, and probably also internal
excitation of the projectile ions. Nevertheless, the various neutral
fragmentation channels are more clearly resolved in the present results. This
demonstrates the wealth of new information accessible by calorimetric detection
of molecular fragments with high relative energy resolution.

\section{Summary and Outlook}\label{sec:sumout}
We have studied the energy resolution function of an MMC detector for kinetic
energy measurements of atomic and molecular projectiles in the energy range from
$\sim 13$ to $150$~keV. For atomic projectiles we have demonstrated that
relative
resolutions down to $\Delta E/E\sim 10^{-3}$ can be achieved. For C and H, the
most relevant atomic projectiles in organic chemistry, relative resolutions
below $10^{-2}$ were reached at all projectile energies studied. Additionally,
we have studied the response of the MMC to molecular projectiles and
demonstrated the capability of calorimetric mass spectrometry to identify
molecular fragmentation products.

We have also created a model of the micro-calorimeter energy resolution for
molecular projectiles. The model was used to fit experimental data from CH$_3^+$
and a very good qualitative and quantitative reproduction of the measured
spectra was found. The quality of the model allowed us to resolve the
contributions of the various energy-loss processes to the detector
energy-response function. For molecules, the dominant distortion of the spectra
originates from energy loss due to backscattering. Based on this finding we have
experimentally demonstrated the strong influence of the absorber material on 
the detector energy-response function. Specifically, using a thin aluminum
coating on the absorber resulted in a strong reduction of the distortion from
backscattering. The validity of our model has been demonstrated for an MMC
detector. Nevertheless, similar energy-loss processes are expected to occur in
all types of cryo-detectors, including other micro-calorimeter types and STJs.
Thus our findings on the MMC energy-response function are likely to be relevant
for cryo-detectors in general.

To the best of our knowledge, our work is the first experimental study of the
energy-response function of cryo-calorimeter detectors for atomic and
molecular projectiles over the  energy range covered. 
This model, which we have developed to estimate the energy-response function,
can be used by us and by others to extrapolate our findings to lower projectile
energies and more complex molecular projectiles. 
In future studies we will investigate other coating
materials in order to further improve the detector energy-response function.

Implementation of micro-calorimeter detectors for fragmentation studies, such
as the DR measurements at CSR proposed in Section~\ref{sec:intro}, requires not
only a high energy resolution, but also a large sensitive area, coincident
fragment detection capability, and position resolution. To meet these
requirements we have taken two different segmented MMC-design approaches.
Segmentation provides the ability to detect coincident particles. The first
approach consists of using large-area detector segments operated independently.
Using a special configuration of thermal links between the absorber, sensor, and
heat bath allows one to determine the incident position within each segment.
This configuration has been realized and successfully tested on a circular
prototype detector of 34~mm in diameter with 16 pie-like segments
\cite{Kampkotter:dipl:2010,Gamer:dipl:2013}. The second approach involves our
development of a large 4~kilo-pixel MMC detector with an active area of
$44.8\times44.8$~mm$^2$. The $64\times64$  particle absorbing pixels making up
this detector are coupled to a square array of paramagnetic temperature sensors.
The sensor pickup coils are wired to provide for each row and each column of
this array a summed temperature signal from all of the sensors in the respective
row or column. This approach reduces the number of readout channels from
$O(N^2)$ to $O(2N)$, where $N$ is the dimension of the square detector. Compared
to the first approach, this pixelized detector will have a more uniform spatial
resolution and higher multi-hit capability \cite{Schulz:master:2015}.

Based on the results of our present MMC energy-response study and of the
large-area position-sensitive MMC developments, briefly described above, we
conclude that MMC detectors are now a realistic option for applications in
atomic and molecular collision experiments and in mass spectrometry at keV
energies. Our priority is to implement such a large detector in the recently
commissioned electrostatic Cryogenic Storage Ring (CSR) at the Max Planck
Institute for Nuclear Physics in Heidelberg, Germany
\cite{Von_Hahn:NIMB:2011,Krantz:JPCS:2011}.  The system will be used for
electron-ion merged-beams studies as explained in Section~\ref{sec:intro}. The
$<10$~K cryogenic environment of CSR facilitates the technical implementation of
the large-area micro-calorimeter. The MMC energy resolution demonstrated here
will be sufficient to efficiently detect and assign neutral products for DR of
molecular ions stored in CSR. We expect achieve mass resolutions of 1 u
for neutral DR products from parent molecular ions of mass $\sim
100$~u and perhaps even heavier. The planned detector segmentation approach will
allow for differentiation of the various DR fragments and their corresponding
masses, thereby enabling the assignment of the various DR fragmentation channels
(see Equation~\ref{eq:DR}) \cite{Buhr:PRA:2010b,Novotny:JPCA:2010}. Moreover,
the position sensitivity connected with the detector segmentation will give us a
measure on the kinetic energy released in the DR reaction for the various
fragmentation channels. As a result, we will also be able to study the internal
excitation of the product atoms and molecules can be investigated
\cite{Zajfman:PRL:1995}. 

The micro-calorimeter detector, in combination with the CSR facility, will be a
powerful tool for detailed studies of electron-capture reactions leading to
neutral products (such as DR and ECD) in unprecedented detail, particularly for
complex molecules. Moreover, we propose to extend molecular ion studies using
micro-calorimeters to other reactions yielding neutral products, such as
photodissociation, dissociative photodetachment, and collision-induced
dissociation. While experimental techniques for studying some of these reactions
exist (e.g., \cite{Lepere:JCP:2007, Pedersen:PRL:2007,Poad:JPCA:2013}), they are
usually unable to reliably distinguish and mass-assign multiple neutral
fragments. The use of a segmented kinetic-energy-sensitive micro-calorimeter can
overcome these limitations also for multi-channel fragmentation reactions
involving complex molecular ions, including species of biochemical interest.

\begin{acknowledgments}
We acknowledge financial support by the Max-Planck-Society. We thank the MPIK
accelerator crew for their excellent support. O.N. and
D.W.S. were supported in part by the NSF Division of Astronomical Sciences 
Astronomy  and  Astrophysics  Grants  program  and by the NASA Astronomy and
Physics Research and Analysis Program.  The detector development at Heidelberg
University was in part financed through the BMBF grant 05P12VHFA5. D.S. 
acknowledges  the  support  of 
the  Weizmann Institute of Science through the Joseph Meyerhoff program. The
work is supported in part by the German-Israeli Foundation for Scientific
Research (GIF under contract no. I-900-231.7/ 2005). We thank Thomas St\"ohlker
and collaborators for letting us use indispensable cryogenic equipment during
our measurements. 
\end{acknowledgments}


\nocite{*}
\bibliography{MMCNovotny}

\end{document}